# Label-free super-resolution color flow imaging using ultrasound phase microscopy

Zhengchang Kou[1†*], Junhang Zhang[2†], Chen Gong[2,3†], Jie Ji[2], Nathiya Vaithiyalingam Chandra Sekaran[1], Zikai Wang[2], Rita J. Miller[1,4], Yaoheng Yang[2], Daniel Adolfo Llano[1,5,6], Qifa Zhou[2,7*] and Michael L. Oelze[1,4,6*]

[1] Beckman Institute for Advanced Science and Technology, University of Illinois at Urbana-Champaign, IL 61801, USA

[2] Alfred E. Mann Department of Biomedical Engineering, University of Southern California, Los Angeles, CA 90089, USA

[3] F.M. Kirby Neurobiology Center, Boston Children's Hospital, and Department of Neurology, Harvard Medical School, Boston, MA 02115, USA

[4] Department of Electrical and Computer Engineering, University of Illinois at Urbana-Champaign, IL 61801, USA

[5] Department of Molecular and Integrative Physiology, University of Illinois at Urbana-Champaign, IL 61820, USA

[6] Carle Illinois College of Medicine, University of Illinois at Urbana-Champaign, IL 61801, USA

[7] Roski Eye Institute, Keck School of Medicine, University of Southern California, Los Angeles, CA 90089, USA

[†] These authors contributed equally: Zhengchang Kou, Junhang Zhang, Chen Gong

[*] Corresponding author: Michael L. Oelze, Qifa Zhou, Zhengchang Kou

## Abstract

Ultrasound vascular imaging is limited by acoustic diffraction, restricting visualization of microvessels essential for understanding organ function and disease. Label-free super-resolution methods exploiting endogenous red blood cells have faced challenges in acquisition time and complexity. Here we introduce ultrasound phase microscopy (UPM), a label-free technique that achieves sub-wavelength resolution flow imaging by exploiting phase differences between consecutively beamformed frames with mismatched apodizations, without requiring localization or tracking. Validated in vivo across multiple species, organs, and ultrasound platforms, UPM attains spatial resolutions better than 5 µm—up to tenfold improvement over conventional color flow imaging—while accelerating data acquisition by nearly two orders of magnitude compared to ultrasound localization microscopy. UPM enables rapid, high-resolution vascular imaging and offers a practical approach for label-free super-resolution vascular imaging.

## Introduction

Ultrasound vascular imaging, including power Doppler imaging (PDI) and color flow imaging (CFI), is widely used in clinical practice. However, its spatial resolution is fundamentally constrained by acoustic diffraction, limiting the visualization of microvessels smaller than roughly one wavelength. This limitation remains a major barrier to studying fine vascular structures essential for brain function, organ perfusion, and disease progression.

Over the past decade, ultrasound localization microscopy (ULM) has transformed vascular imaging by localizing and tracking microbubbles within the vasculature, achieving sub-10 µm resolution beyond the diffraction barrier (*1*). ULM has enabled unprecedented mapping of microcirculation in both two and three dimensions (*2, 3*), facilitating detailed studies of neurovascular dynamics in neuroscience (*4*) and advancing tumor vascular characterization in oncology (*5*). However, its reliance on exogenous contrast agents, long acquisition times, and heavy computational demand limit its use in longitudinal and clinical studies (*6*).

To overcome these constraints, recent efforts have focused on label-free super-resolution approaches that exploit endogenous scatterers, primarily red blood cells (RBCs), instead of microbubbles. Their abundance and ubiquitous distribution make RBCs attractive natural probes for vascular imaging. Several strategies—based on compressed sensing (*7*), localization and tracking (*8, 9*), or deep learning (*10*)—have been proposed. In our previous null-subtraction imaging (*11, 12*) study, envelope signals beamformed with mismatched apodizations were subtracted to enhance the spatial resolution of PDI (*13*) without the extraneous contrast agents. However, these existing label-free methods remain limited by low sensitivity, lengthy acquisitions, inability to resolve flow direction, and complex post-processing. These challenges highlight the need for a generalizable approach to extract super-resolved vascular information from RBCs signals.

Color flow imaging (CFI) is one of the most widely used ultrasound flow imaging modalities that operates with native blood flow. CFI extracts the relative phase offset between frames to reveal flow direction and velocity. However, its spatial resolution is fundamentally limited by the diffraction-defined point spread function, which is determined by the ultrasound frequency, probe geometry, and the beamforming algorithm—typically the delay-and-sum (DAS) method, where signals from each transducer element are delayed and summed to form an image.

To overcome these limitations, we introduce ultrasound phase microscopy (UPM), a label-free super-resolution technique that harnesses the intrinsic acoustic scattering and phase shifts resulting from the flow of RBCs inside vessels. In UPM, we developed an ultrasound spatiotemporal beamforming method that encodes the relative phase offset induced by RBC displacement into the beamwidth. Consequently, UPM achieves phase-resolved super-resolution microvessel imaging purely through beamforming, by reconstructing spatial information from the relative phase offsets. Unlike localization- or tracking-based methods, UPM is computationally inexpensive and supports acquisitions on the order of seconds, making it suitable for dynamic imaging. We validate UPM in vivo across three species, three organ systems, three ultrasound platforms, and four transducers, demonstrating robust performance across frequencies and species. In addition, we explored the capabilities of UPM for studying vasculature changes associated with Alzheimer's disease (AD) and for monitoring dynamic blood flow changes in functional ultrasound (fUS).

The following section details the theoretical framework and practical implementation of UPM, along with in vivo demonstrations.

## Principle of UPM

In UPM, the time-varying backscatter from RBCs flowing within microvessels (Fig. 1A) serves as a direct reporter of blood flow. UPM exploits the relative phase offset generated by the displacement of RBCs between consecutive frames.

During imaging, ultrafast ultrasound plane-wave imaging data (*14*) are acquired using programmable ultrasound imaging research platform within seconds to capture subtle hemodynamic fluctuations. Each frame of the ultrafast ultrasound data comprises multiple plane-wave transmissions at different steering angles (*15*) to improve image quality. Analog-to-digital converters (ADCs) sample and digitize the echo signals from the transducer array to form the raw channel data.

To reconstruct images from the raw channel data, time delays corresponding to the acoustic time of flight are applied to each channel before summation—a process known as DAS beamforming. An apodization window can be applied before summation to suppress sidelobes and mitigate artifacts. Instead of conventional Hanning or Kaiser windows, which emphasize the aperture center, UPM employs three distinct apodizations—ZM, DC1, and DC2—introduced in our previous work (*13*) to generate three sets of beamformed data. Then, a singular-value decomposition (SVD) filter (*13, 16*) is applied to remove stationary tissue clutter while preserving dynamic RBC backscatter.

In conventional CFI, relative phase offsets between consecutive beamformed frames are computed to map blood-flow velocity and direction (*17*). However, the phase response in CFI is inherently dependent on the diffraction-limited beamwidth of the amplitude point-spread function, constraining its lateral resolution. In UPM, four sets of relative phase offsets are computed between two consecutive frames beamformed with mismatched apodizations (Fig. 1B). These phase offsets are then combined into two complementary precursors (PreA and PreB), and the final UPM image is generated by subtracting PreB from PreA. As a result, the beamwidth of the relative phase response varies linearly with axial velocity (Fig. S1), indicating that axial displacements of RBCs coherently modulate the lateral extent of the phase response. This axial-to-lateral modulation encodes sub-wavelength motion information into the phase domain and effectively compresses the lateral main lobe beyond that of the amplitude response. Therefore, UPM achieves super-resolved flow contrast purely through beamforming with three apodizations and relative phase calculation, without any localization or tracking.

Experimental results confirm these theoretical predictions. The UPM and CFI images of a label-free mouse brain acquired at 50-MHz ultrasound are shown in Fig. 1C, with the same hemisphere displayed symmetrically; comparison of both hemispheres is provided in Fig. S2. A manually selected single vessel profile comparison between UPM and CFI is shown in Fig. 1D. A single vessel profile along each processing step of UPM is provided in Fig. S3. A field of view (FOV) of 9.4 x 32.5 x 4.7 $mm^3$ was achieved in this setup. 65 slices were acquired where 4,576 frames (4.76 seconds) ultrafast ultrasound data were acquired in each slice. The vessel radius was locally measured with auto-skeletonization tools. The distributions of vessel radius of both UPM and CFI are shown in Fig. 1E. The average vessel radius measured with UPM and CFI are 8.4 µm and 36.6 µm. The ratio of the measured vessel radius with CFI and UPM at each vessel segment was also calculated and shown in Fig. 1F. The average ratio is 4.6 and the maximum ratio is 10.5. Beyond local measurement, a global resolution evaluation method, Fourier ring correlation (FRC), which is widely used in ULM (*18*), was adopted to evaluate the global resolution performance of UPM. For UPM, the FRC that provides the best performance over all the slices is shown in Fig. 1G. A spatial resolution of 4.8 µm is measured with ½-bit threshold. The measured global spatial resolution is only 15.6% of the acoustic wavelength.

UPM was further validated in mouse brain imaging using three different combinations of ultrasound imaging arrays and research platforms.

In the first case, a 40-MHz ultrasound array was used to provide a larger FOV of 10.2 x 39.5 x 6.3 $mm^3$. 79 slices were acquired where 4,096 frames (4 seconds) ultrafast ultrasound data were acquired in each slice. UPM and CFI images are shown in Fig. S4A and Fig. S4B along with the zoom-in versions in Fig. S4C and Fig. S4D. The distribution of vessel radius comparison between UPM and CFI is shown in Fig. S4E. The average vessel radius measured with UPM and CFI are 9.6 µm and 42.2 µm. The distribution of vessel radius ratio is shown in Fig. S4F. The average vessel radius ratio between CFI and UPM is 4.6 and the maximum ratio is 12.5. The lateral profile comparison between UPM and CFI of a selected vessel is shown in Fig. S4G. The FRC that provides the best performance over all the slices is shown in Fig. S4H. A global spatial resolution of 7.6 µm is measured with ½-bit threshold. The measured global spatial resolution is only 20% of the acoustic wavelength.

In the second case, a 30-MHz ultrasound array was used to provide a FOV of 10.2 x 38.4 x 6.3 $mm^3$. 64 slices were acquired where 4,096 frames (4 seconds) ultrafast ultrasound data were acquired in each slice. UPM and CFI images are shown in Fig. S5A and Fig. S5B. Two magnified comparisons are shown in Fig. S5C and Fig. S5D. The FRC that provides the best performance over all the slices is shown in Fig. S5I. A spatial resolution up to 10.9 µm is measured with ½-bit threshold. The measured global spatial resolution is only 22% of the acoustic wavelength.

In the third case, the previous generation of the ultrasound research platform that has lower sampling frequency and shorter buffer size is used with the same 30-MHz ultrasound array to provide a FOV of 6.8 x 72 x 5.5 $mm^3$. 120 slices were acquired where 1,920 frames (2.4 seconds) ultrafast ultrasound data were acquired in each slice. The corresponding images are shown in Fig. S5E-F. As Fig. S5J shows, a spatial resolution up to 15.4 µm is measured with FRC and ½-bit threshold. The statistics of UPM's spatial resolution in mouse brain imaging across all the slices under different scanning setups are shown in Fig. S5K.

## Generalization of UPM

Instead of limiting the validation of UPM on mouse brain, we have also investigated UPM's performance on larger animal models including rats and rabbits to demonstrate the generalization of UPM.

For rat brain imaging, 30-MHz ultrasound was used with a FOV of 13.6 x 37.2 x 6.3 $mm^3$. 62 slices were acquired where 1,800 frames (2 seconds) ultrafast ultrasound data were acquired in each slice. The UPM and CFI images of rat brain with magnified comparison are shown in Fig. 2A. The FRC that provides the best performance over all the slices is shown in Fig. 2D. In UPM images, a spatial resolution up to 10.1 µm is measured from the FRC with ½-bit threshold. The measured global spatial resolution is only 20% of the acoustic wavelength.

For rat spine imaging, the same scanning system as rat brain imaging was used with a shallower FOV of 13.6 x 30.6 x 4.7 $mm^3$. Two segments of the rat spine were scanned separately due the probe width limitation. 51 slices were acquired. The corresponding UPM and CFI images are shown in Fig. 2B. For the first segment, a spatial resolution up to 6.9 µm was measured from the FRC shown in Fig. 2E with ½-bit threshold. For the second segment, a spatial resolution up to 7.5 µm was measured from the FRC shown in Fig. 2F with ½-bit threshold. The measured global spatial resolution is only 14% of the acoustic wavelength.

To explore the feasibility of UPM in a clinically relevant context, we extended the imaging from rodents' brain to rabbit kidney. Instead of fixing the probe onto a linear stage, the rabbit kidney scan was performed using a handheld probe. To mimic an actual clinical scan, the rabbit was scanned without any breath or cardiac gating, and no motion correction was applied. Therefore, we minimized the acquisition duration to suppress motion artifacts, achieving an acquisition time of only 160 ms (300 frames). To boost the SNR and penetration depth, we used pulse inversion transmission for the rabbit kidney scan. A pair of 10-MHz ultrasound pulses with opposite polarity were transmitted, and the corresponding second harmonic echoes at 20 MHz were accumulated within the Verasonics Vantage hardware to minimize the data size transferred to the host computer. A FOV of 32 x 20.5 mm^2^ was achieved.

UPM and CFI images with pulse inversion ultrasound of rabbit kidney are shown in Fig. 2C. The FRC of UPM is shown in Fig. 2G. A global spatial resolution of 32.7 µm is measured with ½-bit threshold. The measured global spatial resolution is only 42% of the acoustic wavelength.

## Investigate vascular impairment caused by Alzheimer’s disease with UPM

Beyond validating UPM's spatial resolution performance across multiple species and organs, we have also explored UPM's capabilities in investigating the cerebrovascular damage caused by AD. The impairment caused by AD in cerebrovascular has been demonstrated by ULM (*19*). Here we applied UPM to a 5xFAD mouse and a wild-type (WT) mouse at the same age to investigate the vascular impairment of AD without using contrast agents.

The UPM images of the WT and AD mouse brain using 30-MHz ultrasound array and Verasonics Vantage system are shown in Fig. 3A-B. For both mice, 41 slices were acquired to provide a FOV of 10.2 x 24.6 x 6.3 $mm^3$ where 1,792 frames (2 seconds) ultrafast ultrasound data were acquired in each slice. Therefore, the data acquisition could be finished in less than 2 minutes. The magnified comparisons around both sides of hippocampus are shown in Fig. 3C-F.

To quantify the impairment of AD on cerebrovascular, the hippocampus region across all the slices were manually segmented using MATLAB Volume Segmenter. The vascularity of upward flow, downward flow, and bidirectional combined flow was analyzed in the segmented slices and the statistics across all the slices are shown in Fig. 3G. Both upward and downward flow of the WT mouse shows higher average vascularity than the AD mouse. No statistically significant difference was observed in either upward or downward vascularity. However, the combined vascularity demonstrates a statistically significant difference ($p < 0.0001$) between AD and WT mice. The vessel radius analysis was also performed in the segmented slices. The radius measurements of all three categories of vessels show statistically significant differences ($p < 0.0001$) between AD and WT mice (Fig. 3H).

## Demonstrate UPM’s dynamic imaging capability with functional ultrasound

In addition to validating UPM's label-free super-resolution vasculature imaging capabilities and its ability to characterize vascular degeneration in the AD model, we extended the application of UPM from static to dynamic imaging by employing UPM as the beamforming method for fUS (*21*). This approach enables label-free, phase-based functional ultrasound. Unlike conventional fUS, which identifies regions of brain activation by monitoring cerebral blood volume (CBV) changes via a time sequence of PDIs, UPM-based fUS allows dynamic monitoring of flow velocity with super-resolved vessel imaging.

The UPM fUS was validated by imaging light-stimulated brain activation without contrast agents. We repeatedly performed ultrafast ultrasound data acquisition every 1.5 s on a mouse brain during the fUS experiment. Each acquisition contains 640 frames (0.8 second) and results in one UPM image. Each functional trial contains 45 seconds (30 UPM frames) of rest and 15 seconds (10 UPM frames) of light stimulation. Two sets of data were acquired from two coronal planes (0.2 mm distance) of the mouse brain.

The averaged UPM images from two sets of data are shown in Fig. 4A-B overlayed with the corresponding mouse brain coronal schematic view (*22*). The correlation coefficient *r* was calculated between the stimulation function and each pixel in the UPM images sequences. Maps of activated pixels were also overlayed on the top of the UPM images using green color. Activation maps show significant activation in rostro lateral visual area and lateral geniculate nucleus ($r$ = 0.5394 and 0.8076). To demonstrate the transient response, normalized UPM's phase responses extracted from the ROIs marked by red rectangles are plotted along with the stimulation function in Fig. 4C-D.

## Discussion

We present UPM, a label-free, fast acquisition, and fast processing super-resolution microvessel imaging method that can provide spatial resolution better than 5 µm and a FOV larger than 40 mm^2^ with a scanning time shorter than 5 seconds. UPM encodes flow velocity into the beamwidth of the phase response, enabling super-resolution color flow microvessel imaging without localization or tracking of microbubbles or erythrocytes. We also demonstrate UPM's potential in degenerative disease investigation, animal neuroscience research, and clinical translation.

UPM achieved a spatial resolution up to 7.6 µm (Fig. S4) measured globally in a FOV of 10.2 x 6.3 $mm^2$ using 4 seconds of label-free ultrafast ultrasound data acquisition with the same ultrasound probe and ultrasound imaging system as a former study (*20*), in which a spatial resolution of 8.5 µm was achieved with 300 seconds of contrast enhanced ultrafast ultrasound data acquisition. The further validation on different probe and imaging system on the mice brain (Fig. S5) demonstrates UPM's generalizability across different imaging setups.

UPM's potential in clinical translation is demonstrated by the ultra-short label-free handheld acquisition experiment performed on rabbit kidney in which 32 mm wide FOV is achieved with a global spatial resolution up to 32.7 µm. The ultra-short acquisition time (0.16 seconds) enables the feasibility of real-time imaging, with up to 6 frames acquired per second. Because UPM requires far fewer frames than ULM, the resulting ultrasound data size is smaller and processing is faster. The computational cost of UPM is inherently low, as no localization or tracking is required. For the label-free rabbit kidney data, the UPM beamforming time was only 17±0.23 seconds. Although the beamforming time currently exceeds the data acquisition time, future development and optimization of GPU/FPGA beamformers and the use of more parallel accelerators make real-time label-free super-resolution microvessel imaging with UPM highly feasible.

The resolution performance of UPM is determined by the ultrasound frequency, acquisition duration and the performance of acquisition hardware. With current commercially available products, UPM achieves a global spatial resolution better than λ/5 or even λ/6 with an acquisition time two orders shorter than current contrast-enhanced super-resolution ultrasound imaging. With the previous generation of ultrasound research platform, UPM still provides a global spatial resolution better than λ/3. With an ultra-short acquisition duration of 0.16 seconds, UPM could still maintain a spatial resolution better than λ/2.

Although UPM is designed for label-free super-resolution, it is also compatible with contrast-enhanced imaging, as microbubble displacement also generates phase offsets. A microbubble trace experiment was performed using the same acquisition settings as the 40-MHz mouse brain imaging, with only 0.5 seconds of data acquisition. The corresponding results are shown in Fig. S6.

The implications of UPM are broad. In neuroscience, UPM could allow chronic monitoring of cortical microcirculation with fast acquisition, supporting studies of neurovascular coupling and disease progression. In nephrology and transplantation, label-free imaging of renal cortical vasculature could provide early markers of graft perfusion and rejection. More generally, UPM may provide a route for functional ultrasound, biomarker discovery, and preclinical drug evaluation.

Despite these advances, several limitations remain. First, UPM relies on high-frequency ultrasound to maximize red blood cell backscatter, which restricts penetration depth and may limit applications in deep organs or poorly perfused tissues. Second, while UPM is relatively robust to motion, phase stability could be challenged in highly mobile targets or in non-gated acquisitions. Third, although UPM substantially reduces acquisition times compared with ULM, it does not yet achieve real-time imaging, and further optimization of hardware and reconstruction pipelines will be required. Finally, clinical translation will demand systematic evaluation across diverse patient populations, anatomical sites, and disease contexts to confirm safety, reproducibility, and diagnostic value.

By addressing these challenges, future work can extend the impact of UPM, establishing it not only as a research tool but also as a clinically integrated modality for label-free super-resolution vascular imaging.

## Acknowledgments

**Funding:** Z.K. is supported by Beckman Institute Postdoctoral Fellowship**,** J.Z., C.G., J.J. and Q.Z., are supported by R01EY032229 and R01EY035084. R.J.M and M.L.O are supported by R01EB036800.

**Author Contribution:** Z.K. conceived the concept of UPM, developed the imaging sequence, wrote the post-processing and analysis code, and wrote the manuscript. J.Z., C.G, J.J., N.V.C.S., Z.W., R.J.M. performed the animal experiment. G.C. rendered the statistical results and figures. M.Y., R.J.M., D.A.L., Q.Z. and M.L.O. provided guidance on the manuscript preparation. Q.Z. and M.L.O oversaw the animal experiment and manuscript preparation.

# Supplementary Materials

## Methods

### Animal Preparation

**Mouse brain imaging (50 MHz and 30 MHz)**: A 2-month-old C57BL/6J mouse was anesthetized with 5% isoflurane for induction and maintained under 2% isoflurane via a nose cone. Anesthesia depth was monitored via pedal reflex and respiratory rate. To prevent corneal dehydration, GenTeal® Tears Lubricant Eye Gel (Alcon, Fort Worth, TX, USA) was applied to both eyes. The animal was positioned on a temperature-controlled heating pad throughout the procedure. The craniotomy was performed with a lateral width of 6 mm and an anterior-posterior extent from bregma to lambda. The exposed brain surface was covered with ultrasound gel to ensure acoustic coupling. All experimental procedures were approved by the Institutional Animal Care and Use Committee at the University of Southern California.

**Mouse brain imaging (40 MHz)**: Mouse anesthesia was induced using ketamine/xylazine anesthetics, and then mice were placed in a stereotaxic frame with a nose cone supplying oxygen for maintenance. Lidocaine (1%) was intradermally injected into the scalp to supplement anesthesia. Ear bars were used to secure the mouse head to the stereotaxic imaging stage. The scalp of the mouse was removed, and a cranial window was opened on the left and right sides of the skull using a rotary Dremel tool, starting at the sagittal suture and moving laterally to expose the lateral expanse of the cerebral cortex. The animal use protocol was approved by the Institutional Animal Care and Use Committee at the University of Illinois at Urbana-Champaign.

**AD/WT mouse imaging**: 6-month-old male 5xFAD and wild-type (WT) C57BL/6J mice were anesthetized with 5% isoflurane for induction and maintained under 2% isoflurane via a nose cone. Anesthesia depth was monitored via pedal reflex and respiratory rate. To prevent corneal dehydration, GenTeal® Tears Lubricant Eye Gel (Alcon, Fort Worth, TX, USA) was applied to both eyes. The animals were positioned on a temperature-controlled heating pad throughout the procedure. The craniotomy was performed with a lateral width of 6 mm and an anterior-posterior extent from bregma to lambda. The exposed brain surface was covered with ultrasound gel to ensure acoustic coupling. All experimental procedures were approved by the Institutional Animal Care and Use Committee at the University of Southern California.

**Mouse brain functional ultrasound**: Prior to the experiment, the animals were dark-adapted for 12 hours. To record light stimulation activities in the visual system, a full-field strobe flash using a Grass Photic stimulator (Grass Instrument Co., W. Warwick, RI, USA) was placed 10 cm away from the eye to deliver visual stimulation. A function generator (AFG 3252C, Tektronix, Beaverton, OR, USA) was utilized to generate a single cycle burst to synchronize the Verasonics system and the flash stimulator. Specifically, the trigger-out port of the Verasonics system was connected to the trigger-in port of the function generator, and the output port of the function generator was connected to the trigger-in port of the Grass Photic stimulator. During the stimulation phase, the Verasonics system generated a trigger signal every 1.5 s, which prompted the function generator to output a 6-Hz square wave, driving the strobe to flash at a frequency of 6 Hz. The visual stimulation paradigm consisted of three repetitive cycles of 45-s rest and 15-s stimulation.

**Rat brain and spine imaging**: A 6-month-old Long-Evans rat was anesthetized with 3% isoflurane for induction and maintained under 2% isoflurane via a nose cone. Anesthesia depth was monitored via pedal reflex and respiratory rate. To prevent corneal dehydration, GenTeal® Tears Lubricant Eye Gel (Alcon, Fort Worth, TX, USA) was applied to both eyes. For brain imaging, the craniotomy extended from 4 mm anterior to bregma through the bregma-to-lambda interval along the anterior–posterior axis, with a lateral width of 12mm. The exposed brain surface was covered with ultrasound gel to ensure acoustic coupling. For spine imaging, a midline incision was made after shaving the dorsal skin, and the paraspinal muscles were removed. A laminectomy spanning vertebral levels L1–L4 was performed to expose the spinal cord. Ultrasound gel was applied, and the transducer was positioned along the longitudinal axis of the spinal cord (in the direction of the spinal nerves). All experimental procedures were approved by the Institutional Animal Care and Use Committee at the University of Southern California.

**Rabbit kidney imaging**: One in vivo trial was performed with a 3.5-month-old female New Zealand White rabbit weighing 5.8 kg (Charles River Laboratories, Wilmington, MA). The animal use protocol was approved by the Institutional Animal Care and Use Committee at the University of Illinois at Urbana-Champaign. Anesthesia was induced with 5% isoflurane and maintained with 2% isoflurane, both via face mask. The level of anesthesia was monitored by pedal reflex and respiratory rate. Ophthalmic ointment was applied bilaterally, and the rabbit was placed on a heating pad to maintain body temperature. The skin over the left kidney was shaved. The left kidney was scanned transabdominally by handheld VisualSonics MS200 probe.

**Ultrafast Ultrasound Data Acquisition**

Ultrafast ultrasound data were acquired using commercialized ultrasound imaging research systems and ultrasound linear arrays without modification in hardware.

**NXT 50-MHz mouse brain imaging**: A Verasonics NXT 256 high-frequency configuration ultrasound imaging research platform that has 256 channels was used with a VisualSonics MS700 linear array that has 256 elements and a center frequency of 50 MHz. All elements (total width = 9.4 mm) of the probe were used. An ultrafast plane-wave imaging sequence (19 titled plane waves, -13.5° to 13.5° in 1.5° step size, pulse-repetition frequency PRF=18,240 Hz) was developed to acquire ultrafast ultrasound data at a compounded frame rate of 960 Hz. A 50-MHz, 1.5 cycle pulse was generated by the ultrasound system as the transmit signal. Each acquisition has 768 16-bit quantization samples in fast time with a sampling frequency of 125 MHz. The ultrasound data were transferred to a host computer (Dell Precision 7960 with Intel Xeon W7-3565X and 512 GB DDR5 RAM) via PCI-Express Gen3 x16 interface and stored to a NVME SSD array (RAID 0 with 4x Samsung 990 Pro 4TB). 4,576 compounded frames (86,944 acquisitions) were acquired for each imaging plane (4.76 seconds acquisition for each imaging plane). A fixed transmit voltage at 20V was used.

The imaging array was fixed on a motorized stage to translate in elevational direction with a step size of 50.8 µm.

**NXT 40-MHz mouse brain imaging**: A Verasonics NXT 256 high-frequency configuration ultrasound imaging research platform that has 256 channels was used with a VisualSonics MS550S linear array that has 256 elements and a center frequency of 40 MHz. The center 192 elements (total width = 10.2 mm) of the probe were used. An ultrafast plane-wave imaging sequence (15 titled plane waves, -14° to 14° in 2° step size, pulse-repetition frequency PRF=15,360 Hz) was developed to acquire ultrafast ultrasound data at a compounded frame rate of 1024 Hz. A 41.67-MHz, 1.5 cycle pulse was generated by the ultrasound system as the transmit signal. Each acquisition has 1024 16-bit quantization samples in fast time with a sampling frequency of 125 MHz. The ultrasound data were transferred to a host computer (Dell Precision 5860 with Intel Xeon W7-2595X and 256 GB DDR5 RAM) via PCI-Express Gen3 x16 interface and stored to a NVME SSD array (RAID 0 with 8x Samsung 990 Pro 2TB). 4,096 compounded frames (61,440 acquisitions) were acquired for each imaging plane (4 seconds acquisition for each imaging plane). The imaging array was fixed on a motorized stage to translate in elevational direction with a step size of 50 µm. A fixed transmit voltage at 25V was used.

**NXT 30-MHz mouse brain imaging**: A Verasonics NXT 256 high-frequency configuration ultrasound imaging research platform that has 256 channels was used with a VisualSonics MS550D linear array that has 256 elements and a center frequency of 30 MHz. The center 192 elements (total width = 10.2 mm) of the probe were used. An ultrafast plane-wave imaging sequence (15 titled plane waves, -14° to 14° in 2° step size, pulse-repetition frequency PRF=15,360 Hz) was developed to acquire ultrafast ultrasound data at a compounded frame rate of 1024 Hz. A 31.25-MHz, 1.5 cycle pulse was generated by the ultrasound system as the transmit signal. Each acquisition has 1024 16-bit quantization samples in fast time with a sampling frequency of 125 MHz. The ultrasound data were transferred to a host computer (Dell Precision 7960 with Intel Xeon W7-3565X and 512 GB DDR5 RAM) via PCI-Express Gen3 x16 interface and stored to a NVME SSD array (RAID 0 with 4x Samsung 990 Pro 4TB). 4,096 compounded frames (61,440 acquisitions) were acquired for each imaging plane (4 seconds acquisition for each imaging plane). A fixed transmit voltage at 25V was used.

The imaging array was fixed on a motorized stage to translate in elevational direction with a step size of 50.8 µm.

**Vantage 30-MHz mouse brain imaging**: A Verasonics Vantage 256 high-frequency configuration ultrasound imaging research platform that has 256 channels was used with a VisualSonics MS550D linear array that has 256 elements and a center frequency of 30 MHz. The center 128 elements (total width = 6.8 mm) of the probe were used. An ultrafast plane-wave imaging sequence (9 titled plane waves, -18° to 18° in 4.5° step size, PRF=14,400 Hz) was developed to acquire ultrafast ultrasound data at a compounded frame rate of 800 Hz. A 31.25-MHz, 1.5 cycle pulse was generated by the ultrasound system as the transmit signal. Each acquisition has 1280 14-bit quantization samples in fast time with a sampling frequency of 125 MHz (interleave sampling mode was used). The ultrasound data were transferred to a host computer (Dell Precision 5860 with Intel Xeon W7-2595X and 512 GB DDR5 RAM) via PCI-Express Gen3 x16 interface and stored to a NVME SSD array (RAID 0 with 4x Samsung 990 Pro 4TB). 1,920 compounded frames (17,280 acquisitions) were acquired for each imaging plane (2.4 seconds acquisition for each imaging plane). The imaging array was fixed on a motorized stage to translate in elevational direction with a step size of 60 µm. A fixed transmit voltage at 25V was used.

**Vantage 30-MHz AD/WT mouse brain imaging**: A Verasonics Vantage 256 high-frequency configuration ultrasound imaging research platform that has 256 channels was used with a VisualSonics MS550D linear array that has 256 elements and a center frequency of 30 MHz. The center 192 elements (total width = 10.2 mm) of the probe were used. An ultrafast plane-wave imaging sequence (9 titled plane waves, -18° to 18° in 4.5° step size, PRF=16,128 Hz) was developed to acquire ultrafast ultrasound data at a compounded frame rate of 896 Hz. A 31.25-MHz, 1.5 cycle pulse was generated by the ultrasound system as the transmit signal. Each acquisition has 1024 14-bit quantization samples in fast time with a sampling frequency of 125 MHz (interleave sampling mode was used). The ultrasound data were transferred to a host computer (Dell Precision 5860 with Intel Xeon W7-2595X and 512 GB DDR5 RAM) via PCI-Express Gen3 x16 interface and stored to a NVME SSD array (RAID 0 with 4x Samsung 990 Pro 4TB). 1,792 compounded frames (16,128 acquisitions) were acquired for each imaging plane (2 seconds acquisition for each imaging plane). The imaging array was fixed on a motorized stage to translate in elevational direction with a step size of 60 µm. A fixed transmit voltage at 25V was used.

**Vantage 30-MHz mouse brain functional imaging**: A Verasonics Vantage 256 high-frequency configuration ultrasound imaging research platform that has 256 channels was used with a VisualSonics MS550D linear array that has 256 elements and a center frequency of 30 MHz. The center 128 elements (total width = 6.8 mm) of the probe were used. An ultrafast plane-wave imaging sequence (9 titled plane waves, -18° to 18° in 4.5° step size, PRF=14,400 Hz) was developed to acquire ultrafast ultrasound data at a compounded frame rate of 800 Hz. A 31.25-MHz, 1.5 cycle pulse was generated by the ultrasound system as the transmit signal. Each acquisition has 1024 14-bit quantization samples in fast time with a sampling frequency of 125 MHz (interleave sampling mode was used). The ultrasound data were transferred to a host computer (Dell Precision 5860 with Intel Xeon W7-2595X and 512 GB DDR5 RAM) via PCI-Express Gen3 x16 interface and stored to a NVME SSD array (RAID 0 with 4x Samsung 990 Pro 4TB). 640 compounded frames (5,760 acquisitions) were acquired for each power Doppler frame. The time interval between two power Doppler frames was 1.5 seconds. A fixed transmit voltage at 25V was used.

**Vantage 30-MHz rat brain and spine imaging**: A Verasonics Vantage 256 high-frequency configuration ultrasound imaging research platform that has 256 channels was used with a VisualSonics MS550D linear array that has 256 elements and a center frequency of 30 MHz. All 256 elements (total width = 13.6 mm) of the probe were used. An ultrafast plane-wave imaging sequence (9 titled plane waves, -4° to 4° in 1° step size, PRF=16,200 Hz) was developed to acquire ultrafast ultrasound data at a compounded frame rate of 900 Hz. A 31.25-MHz, 1 cycle pulse was generated by the ultrasound system as the transmit signal. Each acquisition has 1024 (768 for spine imaging) 14-bit quantization samples in fast time with a sampling frequency of 125 MHz (interleave sampling mode was used). The ultrasound data were transferred to a host computer (HP Z6A G5 with AMD Threadripper Pro 7985WX and 256 GB DDR5 RAM) via PCI-Express Gen3 x16 interface and stored to a NVME SSD (Samsung 9100 Pro 4TB). 1,800 compounded frames (16,200 acquisitions) were acquired for each imaging plane (2 seconds acquisition for each imaging plane). The imaging array was fixed on a motorized stage to translate in elevational direction with a step size of 60 µm. A fixed transmit voltage at 20V was used.

**Vantage rabbit kidney imaging**: A Verasonics Vantage 256 standard-frequency configuration ultrasound imaging research platform that has 256 channels was used with a VisualSonics MS200 linear array that has 256 elements and a center frequency of 15 MHz. All elements of the probe (total width = 32 mm) were used. An ultrafast plane-wave imaging sequence (9 titled plane waves, -4° to 4° in 1° step size, PRF=33,333 Hz) was developed to acquire ultrafast ultrasound data at a compounded frame rate of 1852 Hz. A pair of 10.83-MHz full cycle pulses with opposite polarity were generated by the ultrasound system for each steering angle and the received signals from both pulses were accumulated inside the Verasonics system to implement pulse inversion and reduce the data size. A 10.83-MHz, 1 cycle pulse was generated by the ultrasound system as the transmit signal. Each acquisition has 1664 14-bit quantization samples in fast time with a sampling frequency of 62.5 MH. The ultrasound data were transferred to a host computer (Dell Precision 5820 with Intel Xeon W-2255 and 128 GB DDR4 RAM) via PCI-Express Gen3 x16 interface and stored to a NVME SSD (WD AN1500 1TB). 300 compounded frames (5,400 acquisitions) were acquired for each imaging plane (162 milliseconds acquisition). A fixed transmit voltage at 30V was used.

**Microbubble trace experiment**: As introduced in (*13*), a small number of microbubbles (Lantheus DEFINITY) were used in the microbubble trace experiment, where the microbubbles were pushed away from the transducer face and recorded over time. The same imaging setup used for the 40-MHz mouse brain scan was used. Instead of acquiring 4,096 frames, only 512 frames were acquired in the microbubble trace experiment. A fixed transmit voltage at 6V was used.

**Ultrasound Phase Microscopy**

After the acquisition of ultrafast ultrasound data, the reconstruction of label-free super-resolution microvessel image was performed using the UPM algorithm. The UPM process is composed of three steps, which are clutter filtering, beamforming, and phase subtraction.

The singular value decomposition (SVD) based clutter filter was performed on the raw channel data to filter out the static tissue signal and retain the native blood flow signal.

The clutter filtered ultrasound channel data are delayed and summed with three sets of weights that are also used in the NSI. Three samples are beamformed at each pixel location following the process below:

$$\begin{bmatrix} S_{zm} \\ S_{dc1} \\ S_{dc2} \end{bmatrix} = [\boldsymbol{W}_{zm} \quad \boldsymbol{W}_{dc1} \quad \boldsymbol{W}_{dc2}]^T [\boldsymbol{R}_{delayed} \quad \boldsymbol{R}_{delayed} \quad \boldsymbol{R}_{delayed}] \tag{1}$$

where $S_{zm}$, $S_{dc1}$ and $S_{dc2}$ are the three beamformed samples, $\boldsymbol{W}_{zm}$, $\boldsymbol{W}_{dc1}$ and $\boldsymbol{W}_{dc2}$ are the three weight column vectors used in NSI, and $\boldsymbol{R}_{delayed}$ is a column vector that contains delayed data samples across the subaperture. A fixed f-number of 1 was used to determine the subaperture size for beamforming samples at different depths. The beamformed samples at the same pixel location across different titled angles are summed together for coherent compounding. The same process was repeated for each pixel location and each frame to finish the ultrafast ultrasound beamforming.

The phase subtraction is performed on three sets of beamformed samples following the process below:

$$P_1 = \sum_{n=1}^{nf-1} \arg\left(\check{S}_{dc1}^{n} * \left(-\check{S}_{dc2}^{n+1}\right)^*\right) \tag{2}$$

$$P_2 = \sum_{n=1}^{nf-1} \arg\left(\check{S}_{dc2}^{n} * \left(-\check{S}_{dc1}^{n+1}\right)^*\right) \tag{3}$$

$$P_3 = \sum_{n=1}^{nf-1} \arg\left(\check{S}_{dc1}^{n} * \left(\check{S}_{zm}^{n+1}\right)^*\right) \tag{4}$$

$$P_4 = \sum_{n=1}^{nf-1} \arg\left(\check{S}_{dc2}^{n} * \left(-\check{S}_{zm}^{n+1}\right)^{*}\right) \quad (5)$$

$$P_{PreA} = (P_1 + P_2) \quad (6)$$

$$P_{PreB} = (P_3 + P_4) \quad (7)$$

$$P_{UPM} = P_{PreB} - P_{PreA} \quad (8)$$

Where $nf$ is the number of frames within one data set and $n$ is the frame index.

$P_{PreA}$ and $P_{PreB}$ are the two precursors of $P_{UPM}$. The final images are generated by accumulating $P_{UPM}$ through all the frames acquired on the same imaging plane.

The above process was performed on an ASUS ESC4000A-E12P server (AMD EPYC 9554 64-core processor, 768 GB DDR5 RAM and two NVIDIA RTX 6000 Ada GPUs). Customized UPM beamforming software was built with CUDA and compiled as Matlab MEX functions. Fixed DC offset at 0.32 was selected empirically for the weights used in the mouse brain, rat brain, and rat spine imaging. For rabbit kidney imaging, a fixed DC offset at 0.72 was used.

**Quantitative Measurement**

Skeletonization was performed using Matlab function *bwskel* and the microvessel radius was measured by finding the minimum Euclidean distance between a voxel on the skeletonized vessel to the voxel that has half value of itself. The vessel density was measured by the ratio between the number of voxels occupied by the vessels and the total number of voxels inside the region of interest.

UPM image resolution was estimated with FRC (*18*). The ½-bit threshold was used to evaluate the image resolution.

**Theoretical Analysis**

The complex beam response of the three apodizations can be evaluated using far field approximation as:

$$H_{Apod}(k_x) = \sum_{n=-M}^{M-1} w_{Apod}[n] e^{-jk_x nd} \quad (9)$$

Where $2M = N$ is the number of elements in the ultrasound array, $k_x = \frac{2\pi}{\lambda}\sin(\theta)$ is the lateral wave number, and $d$ is the pitch of the ultrasound transducer array. Evaluating Eq. 9 using three apodizations results in the following equations.

$$H_{zm}(k_x) = -jCB \quad (10)$$

$$H_{dc1}(k_x) = C(A - jB) \quad (11)$$

$$H_{dc2}(k_x) = C(A + jB) \quad (12)$$

$$C = \frac{e^{\frac{jk_x d}{2}}}{\sin\left(\frac{k_x d}{2}\right)} \quad (13)$$

$$A = \gamma \sin\left(\frac{k_x N d}{2}\right) \tag{14}$$

$$B = 2sin^2\left(\frac{k_x N d}{4}\right) \tag{15}$$

Assuming the phase shift between two successive acquisitions caused by the displacement of red blood cells is $\Delta\varphi = \frac{4\pi f_c v \cos\sigma}{c_0 PRF}$, where $f_c$ is the center frequency, $v$ is the blood flow speed, $\sigma$ is the angle between the blood flow direction and the ultrasound beam direction, $c_0$ is the speed of sound. When we reduce $nf$ to 2, $P_1$ to $P_4$ can be expressed as

$$P_1 = \arg\Big(\big(C(A - jB)\big) * C^*(-A + jB) * e^{-j\Delta\varphi}\Big) \tag{16}$$

$$P_2 = \arg\Big(\big(C(A + jB)\big) * C^*(-A - jB) * e^{-j\Delta\varphi}\Big) \tag{17}$$

$$P_3 = \arg\Big(\big(C(A - jB)\big) * jC^*B * e^{-j\Delta\varphi}\Big) \tag{18}$$

$$P_4 = \arg\Big(\big(C(A + jB)\big) * -jC^*B * e^{-j\Delta\varphi}\Big) \tag{19}$$

Then the $P_{UPM}$ can be calculated from Eq.6-8.

Theoretical phase responses for each step at axial velocities of 1, 2, and 3 mm/s are shown in Fig. S1.

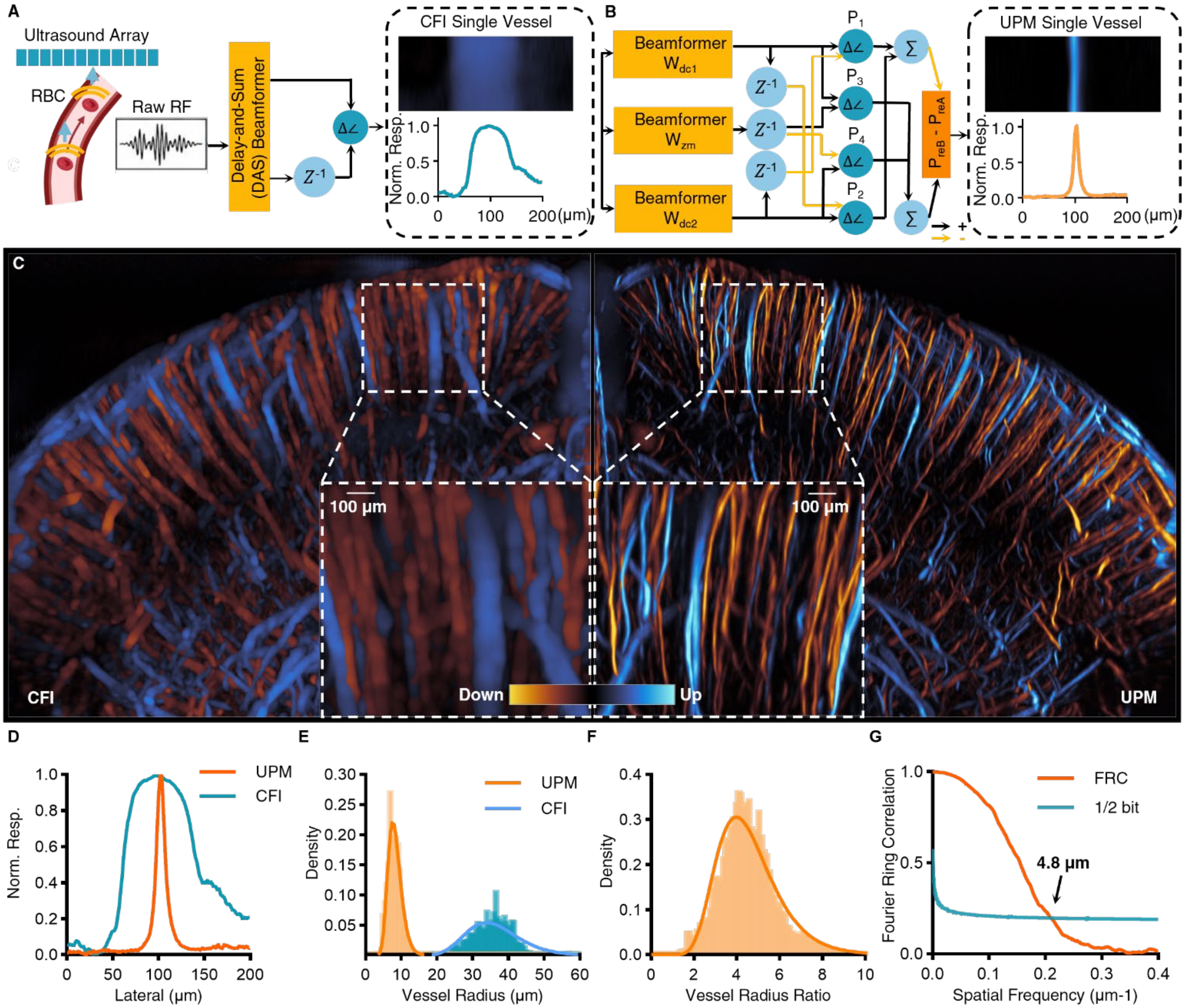

Figure 1. Principle of UPM. (**A**) Schematic of conventional CFI. RBCs flowing through microvessels generate time-varying backscatter signals that serve as intrinsic reporters of blood flow. The axial component of RBCs' motion between consecutive frames produces small phase perturbations in the received echoes. Conventional CFI uses DAS as the beamforming method. (**B**) UPM processing pipeline. Three sets of beamformed images are generated using mismatched apodizations (ZM, DC1, DC2) and paired to compute four relative phase-offset maps between consecutive frames. These phase offsets are then combined into two complementary precursors: PreA and PreB. The final UPM image is obtained by subtracting PreB from PreA. (**C**) Comparison of UPM and CFI in a label-free mouse brain acquired with 50 MHz ultrasound. UPM and CFI images of one side of brain are displayed symmetrically to demonstrate the super-resolution performance of UPM. Full-width comparison is shown in Fig. S2. (**D**) Single vessel lateral profile comparison between UPM and CFI. (**E-F**) Vessel radius distribution comparison between UPM and CFI and resolution improvement ratio of UPM over CFI from selected region of 50 MHz mouse brain dataset. (**G**) FRC of UPM from the 50 MHz mouse brain dataset. A spatial resolution up to 4.8 µm was measured using the ½ bit threshold.

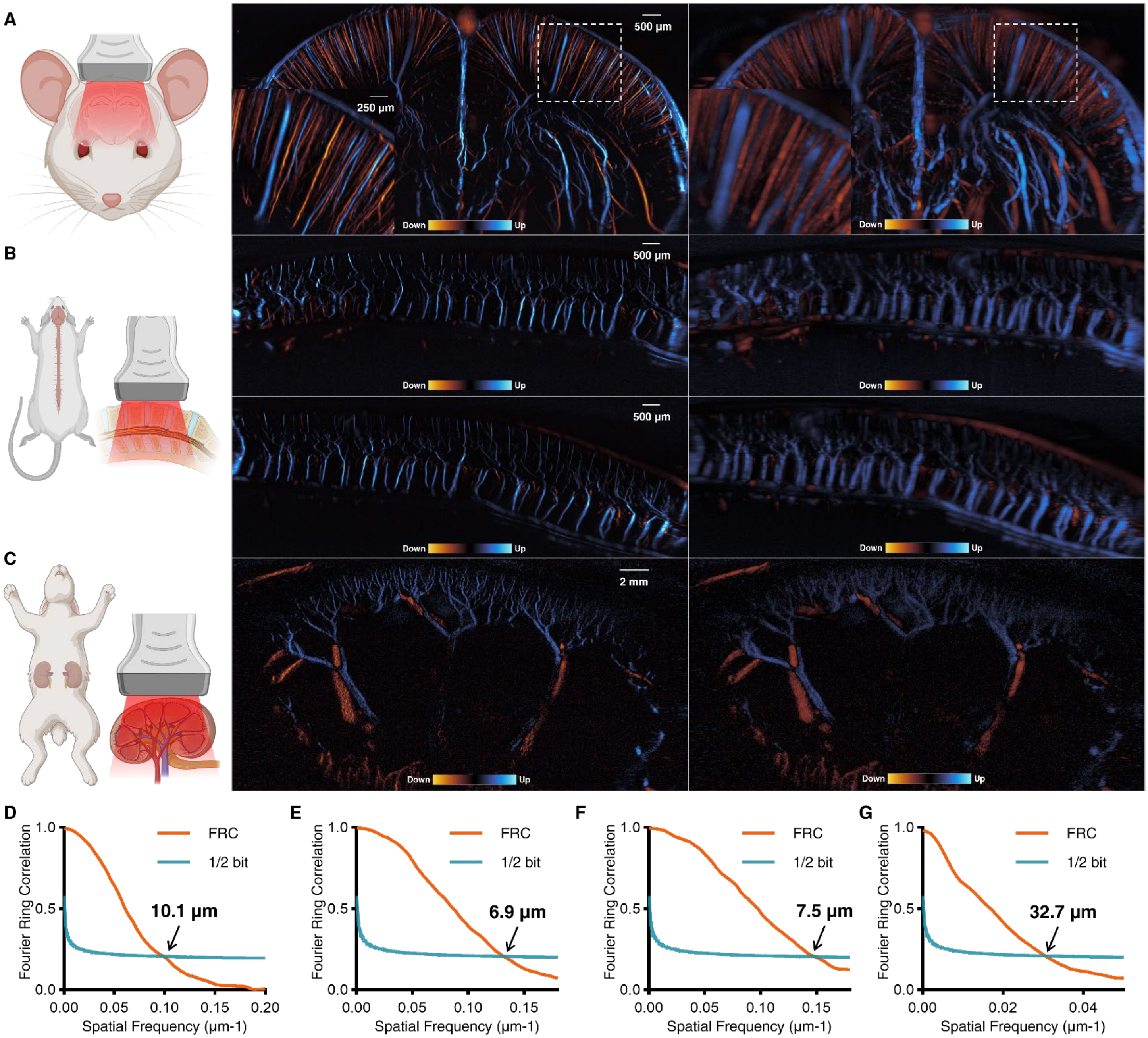

Figure 2. Scalability of UPM in different animal models and ultrasound imaging setups. (**A**) UPM and CFI volume of the rat brain acquired at 30 MHz. (**B**) UPM and CFI volumes of the rat spine acquired at 30 MHz. (**C**) UPM and CFI image of the rabbit kidney acquired with 10 MHz transmission pulse inversion sequence. (**D**) FRC of UPM from the 30 MHz rat brain dataset. A spatial resolution up to 10.1 µm was measured using the ½ bit threshold. (**E-F**) FRCs of UPM from the 30 MHz rat spine dataset. A spatial resolution up to 6.9 µm was measured using the ½ bit threshold. (**G**) FRC of UPM from the rabbit kidney dataset. A spatial resolution up to 32.7 µm was measured using the ½ bit threshold.

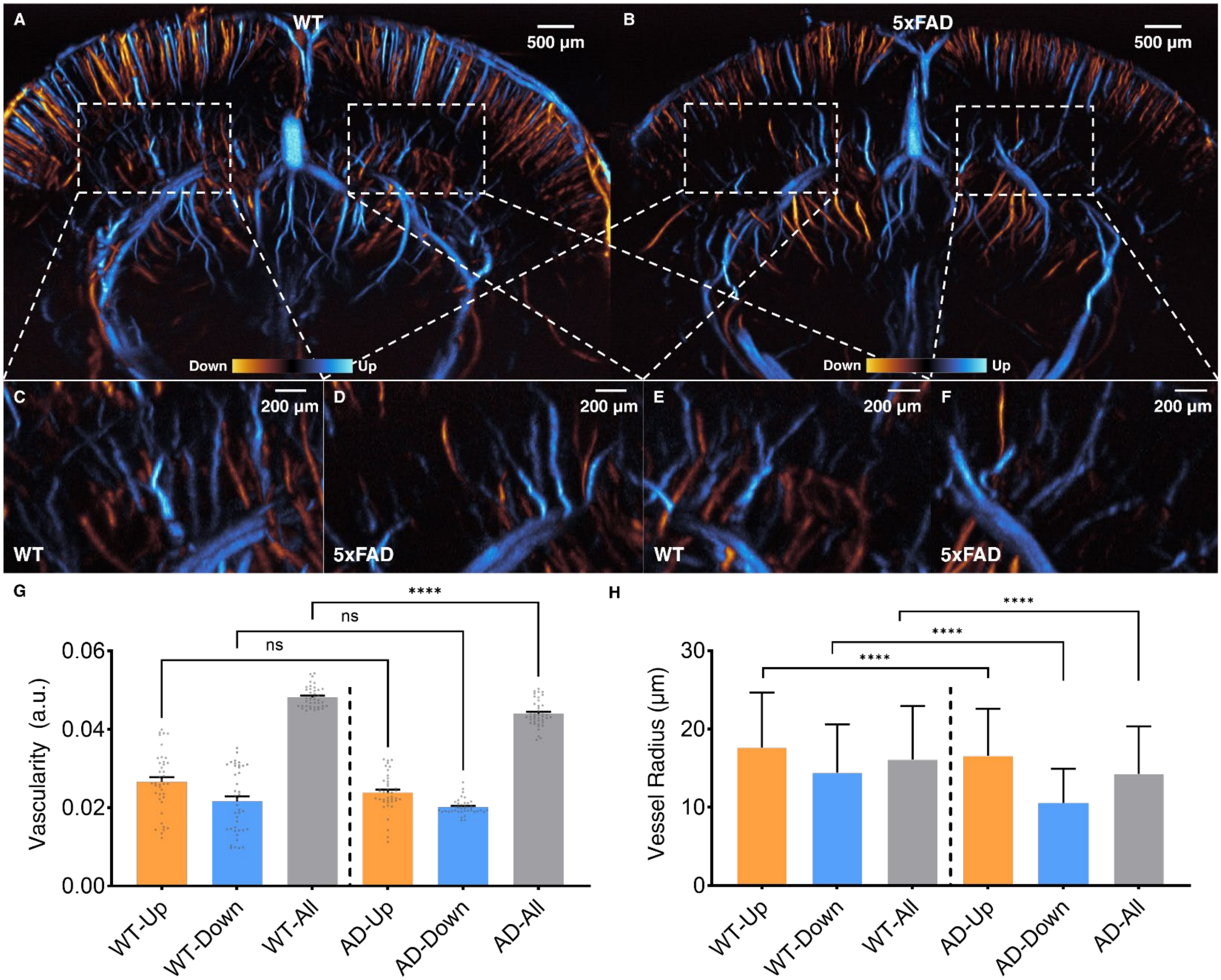

Figure 3. UPM enabled label-free microvasculature analysis of AD's impact on brain. (**A-B**) UPM of WT and 5xFAD mice brain. (**C-F**) Zoomed comparison between WT and 5xFAD mice brain in hippocampus. (**G**) Comparison of the vascularity in the hippocampus between WT and 5xFAD mice brain. (**H**) Comparison of the vessel radius in the hippocampus between WT and 5xFAD mice brain.

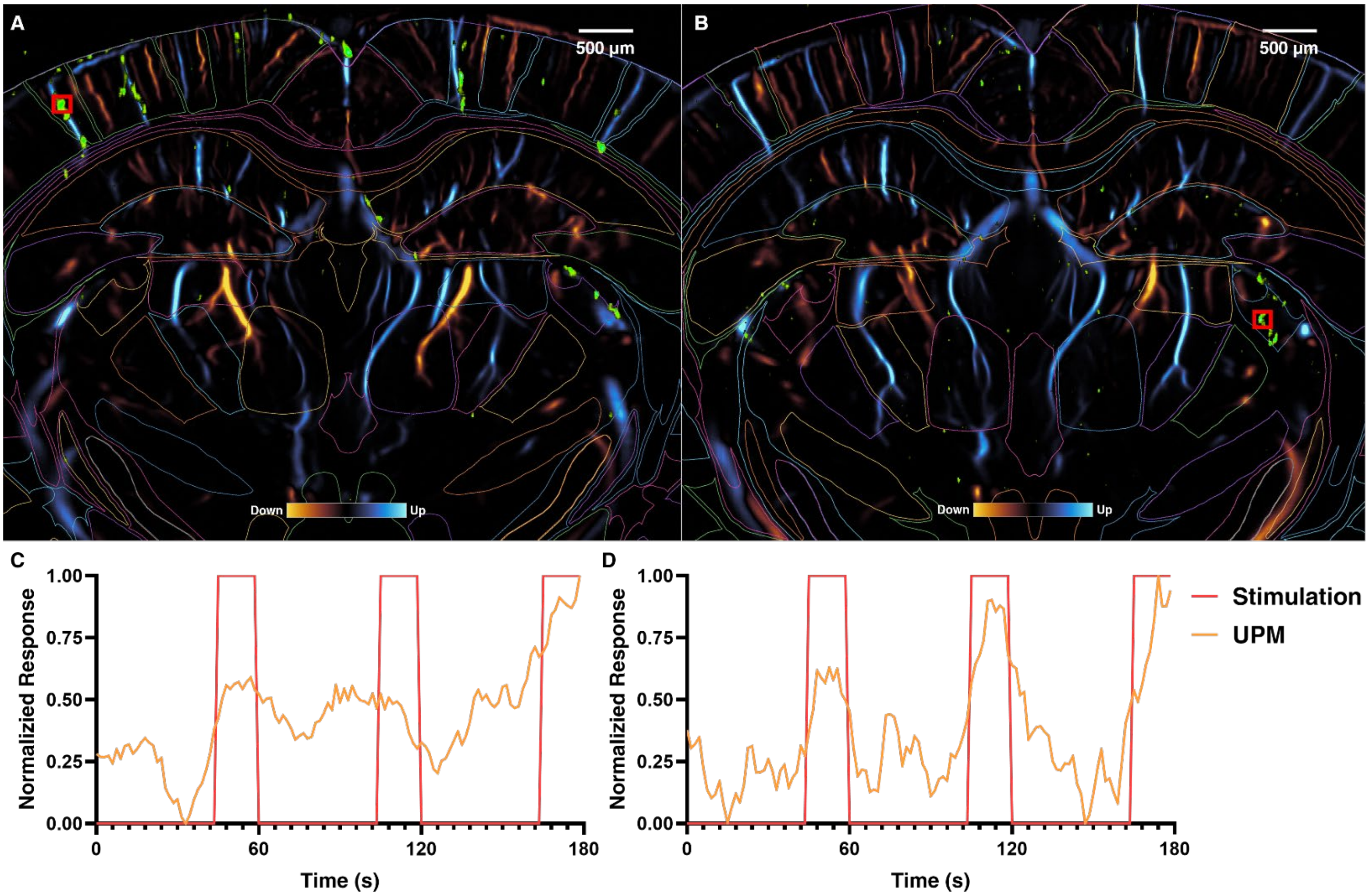

Figure 4. Label-free functional ultrasound imaging with light stimulation using UPM. (**A-B**) UPM images of the mouse brain at two planes that are 0.2 mm separated. Coronal schematic view (color lines) and correlation results (green) are overlayed on top of the UPM images. (**C-D**) Normalized phase response in the ROI marked with red rectangle in (A-B) during the functional ultrasound imaging session composed of 3 strobe light stimulation trials.

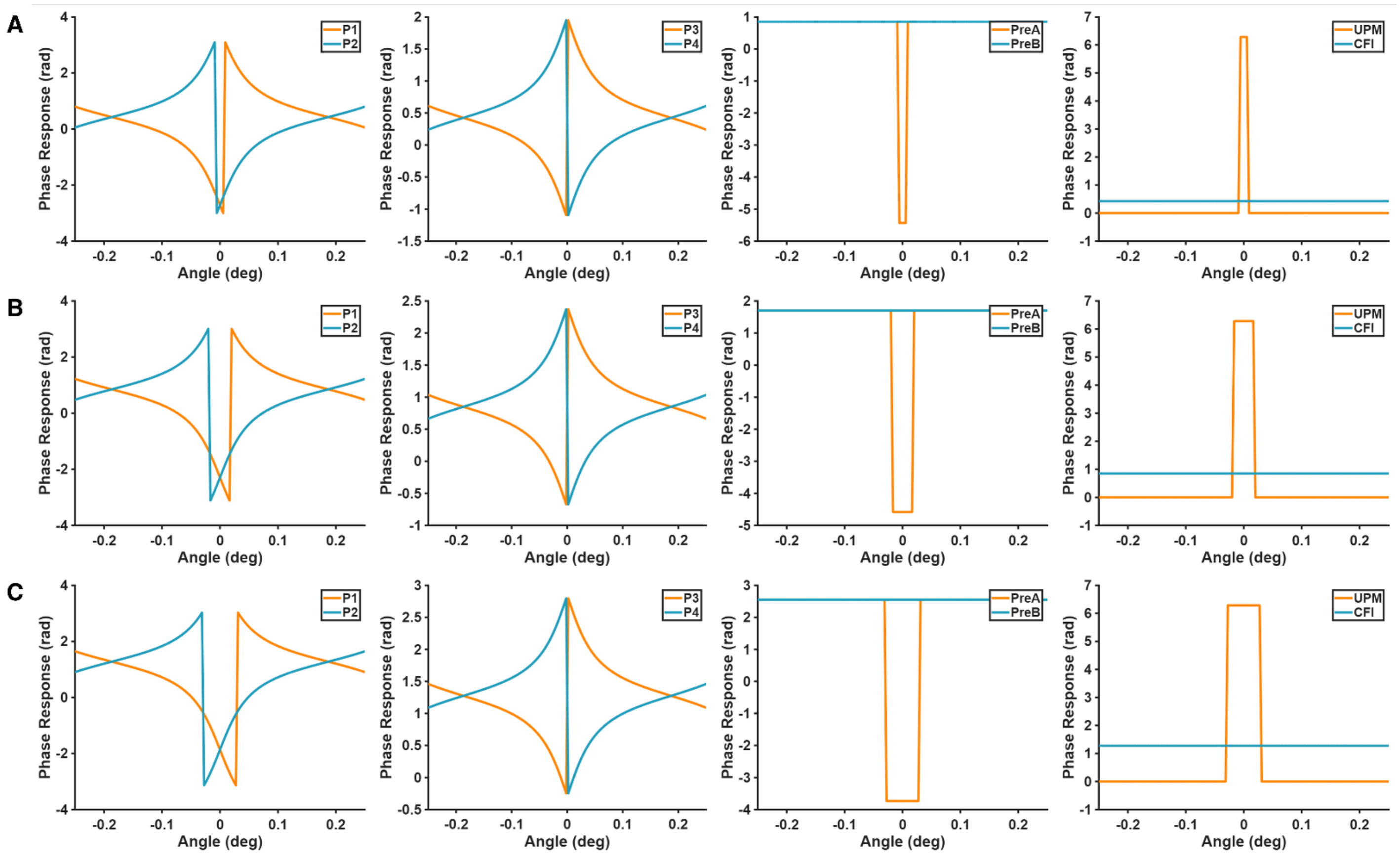

Figure S1. Theoretical phase responses of UPM at different flow speeds. (**A**) Theoretical phase responses of all steps in UPM process with vertical speed at 1mm/s. (**B**) Theoretical phase responses of all steps in UPM process with vertical speed at 2mm/s. (**C**) Theoretical phase responses of all steps in UPM process with vertical speed at 3mm/s.

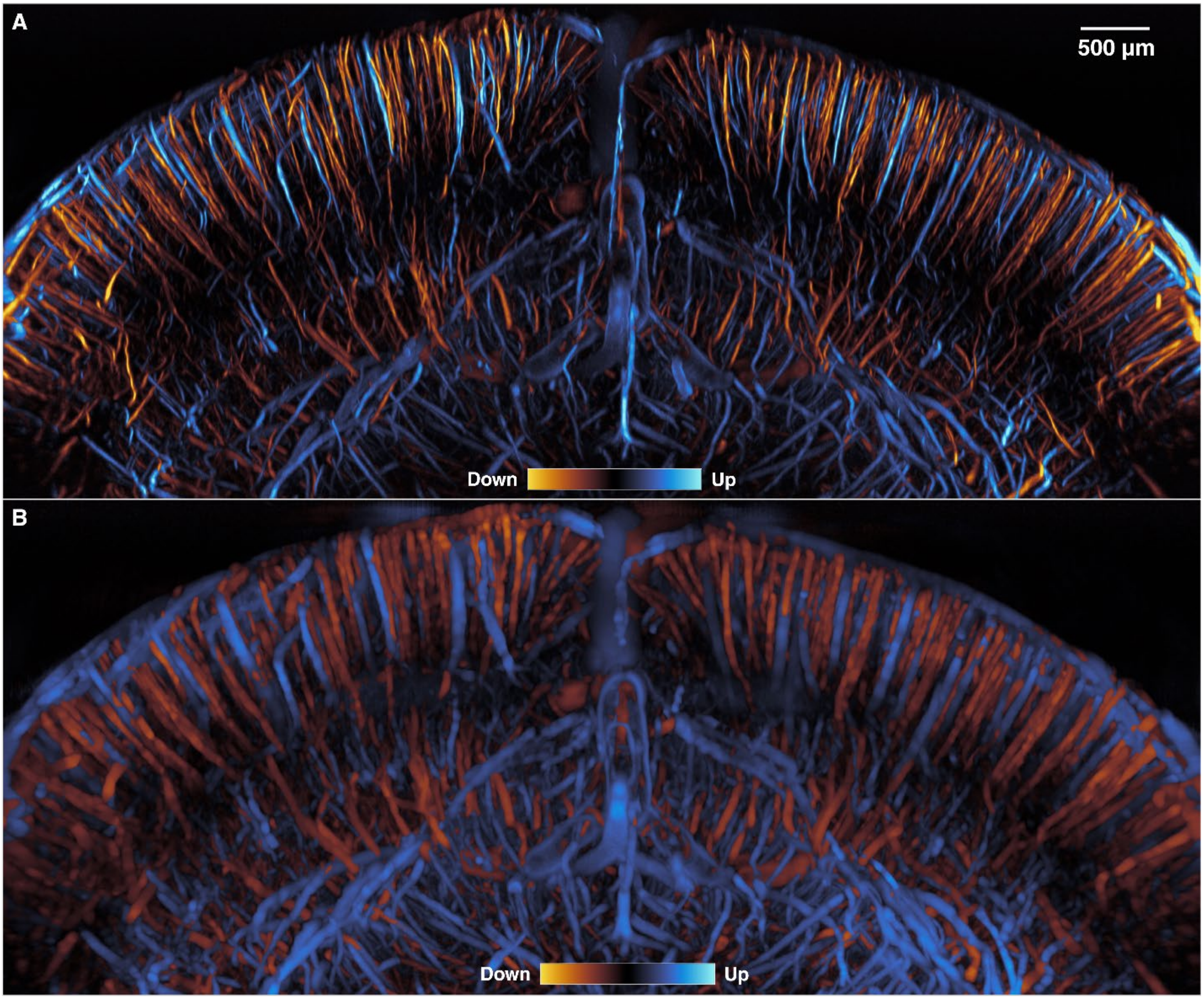

Figure S2. Full comparison between UPM and CFI in a label-free mouse brain (50 MHz). (**A**) UPM volume of the mouse brain acquired at 50 MHz. (**B**) Corresponding CFI volume of the same brain.

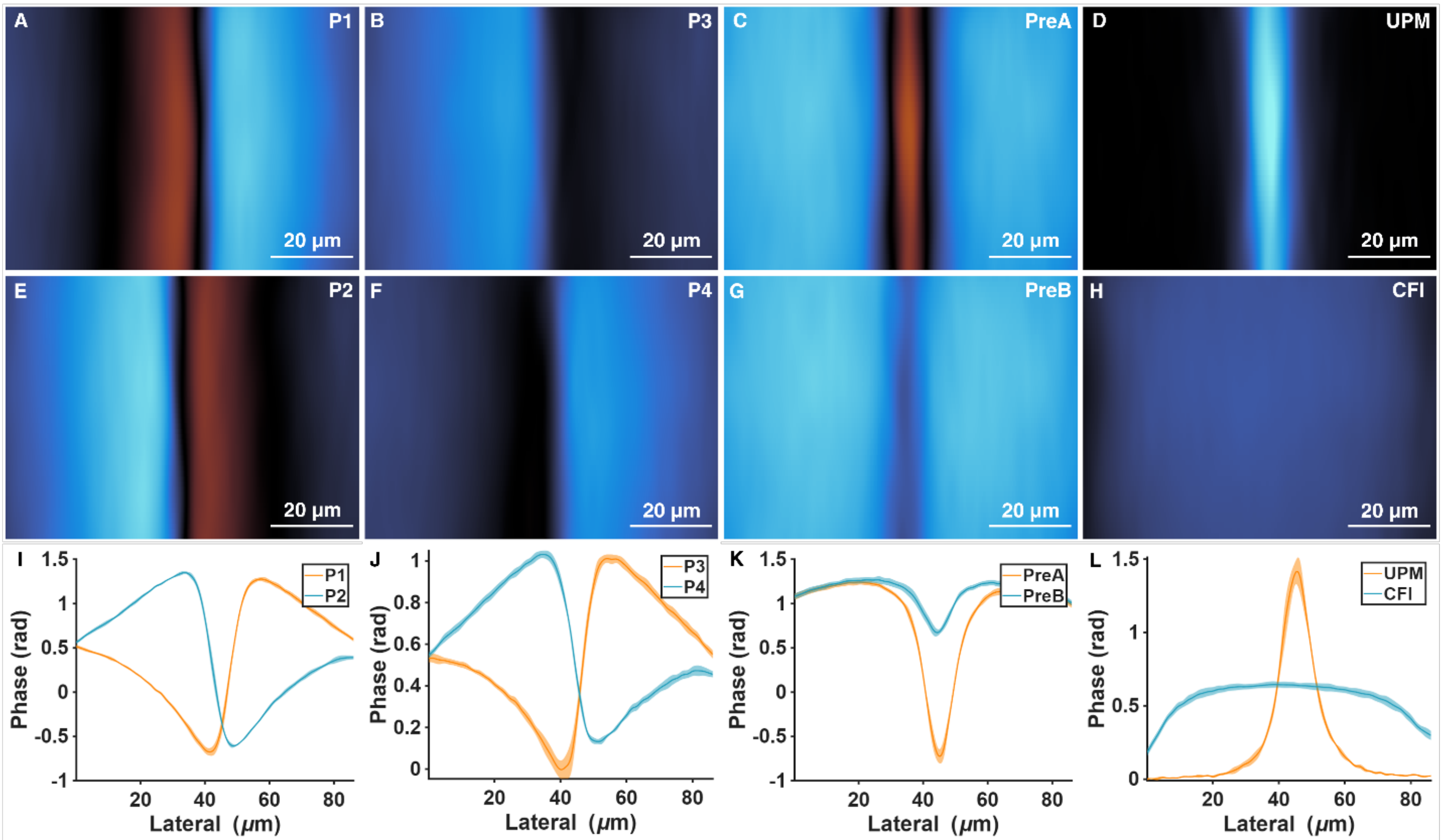

Figure S3. Experimental validation of UPM processing pipeline and intermediate responses from the 50 MHz mouse brain dataset. (**A-H**) Magnified view of a microbubble trace through each step of UPM. (**I-L**) Lateral profile of each step of UPM.

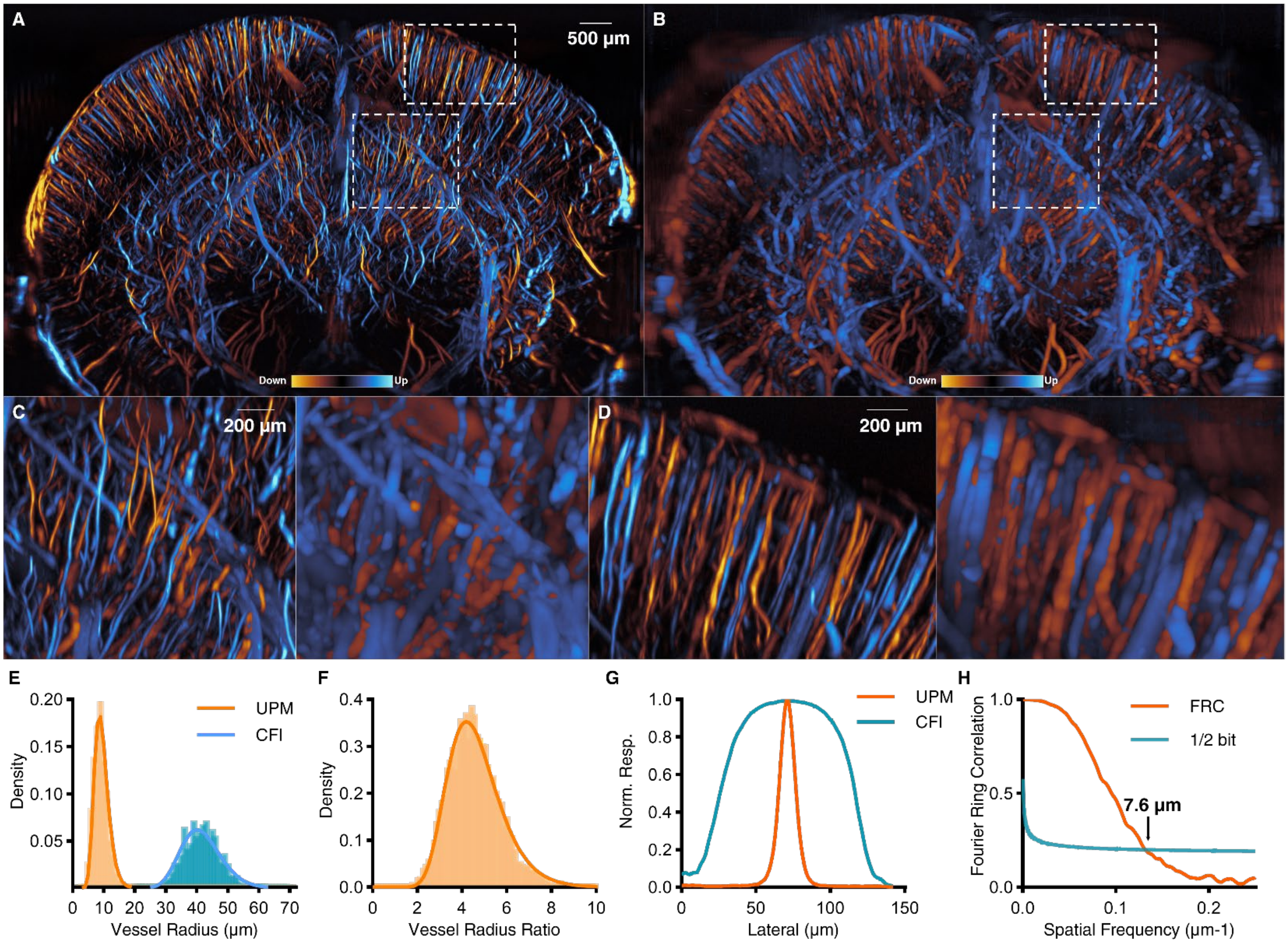

Figure S4. Validation of UPM with 40 MHz ultrasound. (**A**) UPM volume of the mouse brain acquired at 40 MHz. (**B**) Corresponding CFI volume of the same brain. (**C-D**) UPM and CFI comparison in magnified regions from (A). (**E**) FRC of UPM from the 40 MHz mouse brain dataset. A spatial resolution up to 7.6 µm was measured using the ½ bit threshold. (**F-G**) Vessel radius distribution comparison between UPM and CFI and resolution improvement ratio of UPM over CFI from selected region of 40 MHz mouse brain dataset. (**G**) Single vessel cross section comparison between UPM and CFI.

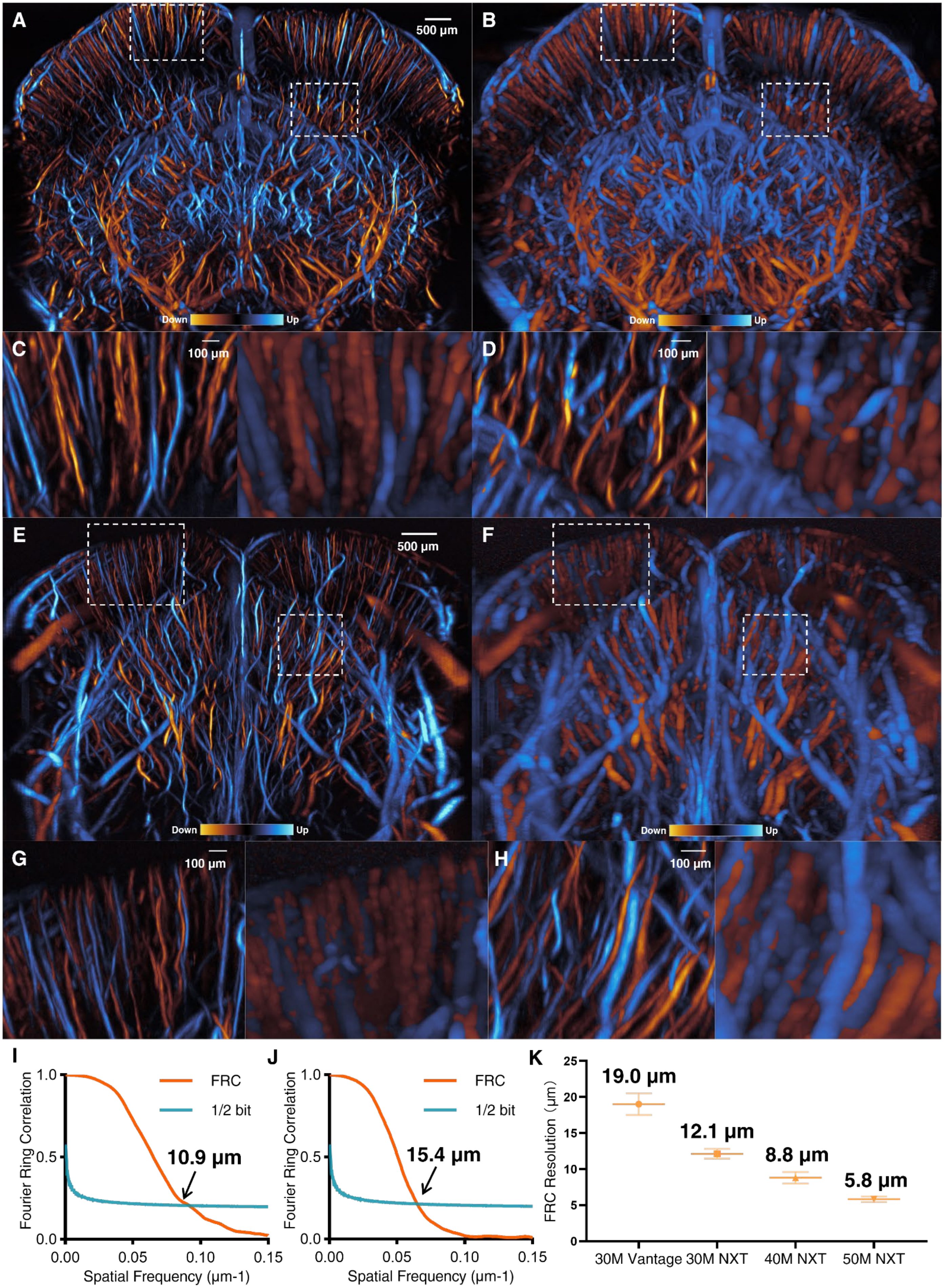

A
500 μm
B
Down
Up
C
100 μm
D
100 μm
E
500 μm
F
G
100 μm
H
100 μm
I
Fourier Ring Correlation
FRC
1/2 bit
10.9 μm
Spatial Frequency (μm-1)
J
15.4 μm
K
FRC Resolution (μm)
19.0 μm
12.1 μm
8.8 μm
5.8 μm
30M Vantage
30M NXT
40M NXT
50M NXT

Figure S5. Validation of UPM with 30 MHz ultrasound and different ultrasound research platforms. (**A**) UPM volume of the mouse brain acquired at 30 MHz with Verasonics NXT. (**B**) Corresponding CFI. (**C-D**) UPM and CFI comparison in magnified regions from (A). (**E**) UPM volume of the mouse brain acquired at 30 MHz with Verasonics Vantage. (**F**) Corresponding CFI. (**G-H**) UPM and CFI comparison in magnified regions from (E). CFI and UPM volume of the mouse brain acquired at 30 MHz. (**I**) FRC of UPM from the 30 MHz Verasonics NXT mouse brain dataset. A spatial resolution up to 10.9 µm was measured using the ½ bit threshold. (**J**) FRC of UPM from the 30 MHz Verasonics NXT mouse brain dataset. A spatial resolution up to 15.4 µm was measured using the ½ bit threshold. (**G**) Statistics of spatial resolution measured with FRC and ½ bit threshold from four datasets shown in Fig.1-3 using both Vantage and NXT systems at different frequencies and acquisition durations.

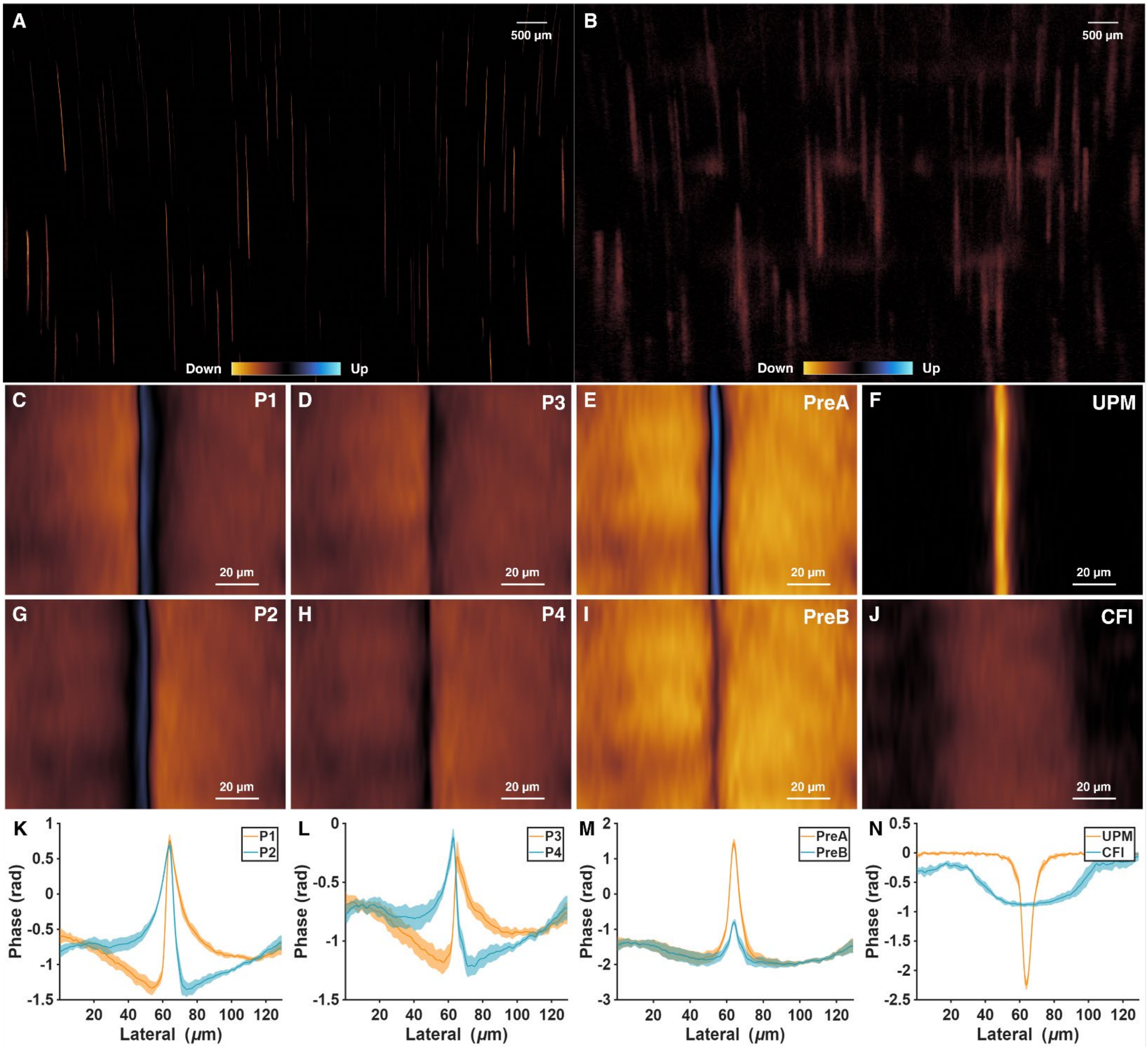

Figure S6. Experimental validation of UPM processing pipeline and intermediate responses from the 40MHz microbubble trace experiment dataset. Microbubbles were pushed downwards by the ultrafast ultrasound imaging sequence and leave traces. (**A**) UPM image of 0.5 seconds microbubble trace data. (**B**) Corresponding CFI image. (**C-J**) Magnified view of a microbubble trace through each step of UPM. (**K-N**) Lateral profile of each step of UPM.